\documentclass[aps,prl,twocolumn,superscriptaddress]{revtex4}

\usepackage{bm}
\usepackage{amsmath}
\usepackage[dvipdfmx]{graphicx}
\usepackage{float}
\usepackage{color}

\usepackage {ulem}

\newcommand{\alphaithree}{{$\alpha$-(BEDT-TTF)$_2$I$_3$\ }}
\newcommand{\nidmdt}{{[Ni(dmdt)$_2$]\ }}
\newcommand{\ToneTini}{{1/$^iT_1T$\ }}
\newcommand{\ToneTbar}{{$\overline{1/T_1T}$\ }}
\newcommand{\vF}{{$v_{\rm F}\ $}}
\newcommand{\Kiso}{{$K_{\rm iso}$\ }}

\begin{document}


\title{NMR verification of Dirac nodal lines in a single-component molecular conductor}


\author{Takahiko Sekine}
\author{Keishi Sunami}
\author{Takumi Hatamura}
\author{Kazuya Miyagawa}
\affiliation{Department of Applied Physics, University of Tokyo, Bunkyo-ku, Tokyo 113-8656, Japan}
\author{Kenta Akimoto}
\author{Biao Zhou}
\affiliation{Department of Chemistry, Nihon University, Setagaya-ku, Tokyo, 156-8550, Japan.}
\author{Shoji Ishibashi}
\affiliation{Research Center for Computational Design of Advanced Functional Materials (CD-FMat), National Institute of Advanced Industrial Science and Technology (AIST), Tsukuba 305-8568, Japan.}
\author{Akiko Kobayashi}
\affiliation{Department of Chemistry, Nihon University, Setagaya-ku, Tokyo, 156-8550, Japan.}
\author{Kazushi Kanoda}
\affiliation{Department of Applied Physics, University of Tokyo, Bunkyo-ku, Tokyo 113-8656, Japan}

\date{\today}

\begin{abstract}
The Dirac nodal line (DNL) is a novel form of massless Dirac fermions that reside along lines in momentum space. Here, we verify the DNLs in the molecular material, [Ni(dmdt)$_2$], with the combined NMR experiments and numerical simulations. The NMR spectral shift and spin-lattice relaxation rate divided by temperature, 1/$T_1 T$, decrease linearly and quadratically with temperature, respectively, and become constant at low temperatures, consistent with slightly dispersive DNLs with small Fermi pockets. Comparison of these results with model simulations of DNLs reveals the suppression of the Fermi velocity and the enhancement of antiferromagnetic fluctuations due to electron correlation as well as the influence of the Landau quantization. The present study offers a demonstration to identify the DNL and evaluate the correlation effect with NMR.
\end{abstract}


\maketitle

Linear energy-momentum dispersions that cross at a single point constitute a band structure of vertex-shared cones, dubbed Dirac cones. Excitations around the vertex (Dirac point) are described by pseudorelativistic massless quasiparticles, called massless Dirac fermions (DFs), which have extraordinary properties arising from their massless and topological natures in conjunction with mutual interactions as revealed in monolayer graphene \cite{Novoselov2004, Novoselov2005,CastroNeto2009,Kotov2012} and an organic layered crystal, \alphaithree\cite{Tajima2006,Katayama2006,Tajima2009,Hirata2017, Hirata2021}. Dirac nodal line (DNL) semimetals, where the Dirac points are continuously connected in lines in $k$ space, have recently attracted considerable interest as a novel form of DFs. The emergence of DNLs has been theoretically and experimentally suggested in several systems such as ZrSiX (X = Se, Te) \cite{Tian2021,Fu2019,Topp2016}, ZrX$_2$ (X = Se, Te) \cite{Tian2020,Kar2020}, AlB$_2$ \cite{Takane2018}, CaAgX (X = P, As) \cite{Hirose2020,Yamakage2016,Takane2018a}, and [Pd(dddt)$_2$] \cite{Kato2017, Kato2017a}. The single-component molecular conductor \nidmdt (Fig. 1(a)) is also theoretically suggested to host DNLs \cite{Zhou2019, Kobayashi2021}. The first-principles calculations predict tilted 2D Dirac cones in the $k_b$--$k_c$  plane, whose Dirac points form two dispersed DNLs along $k_a$ with the width of 29 meV around $E_\mathrm{F}$, leading to small hole-like and electron-like Fermi pockets (Fig. 1(b)). \nidmdt is the first ambient-pressure organic DF candidate material notably having DNLs crossing $E_\mathrm{F}$ without any other bands. Considering intricate situations in other systems (coexisting other metallic bands in ZrSiX and ZrX$_2$, DNLs located apart from $E_\mathrm{F}$ due to defects and impurities in CaAgX, large dispersion of DNLs in AlB$_2$, and DNLs emerging only at high pressures above 12.6 GPa in [Pd(dddt)$_2$]), the present system can be a unique model material for studying the pure DNLs.

\begin{figure}
\centering
\includegraphics[width = \columnwidth]{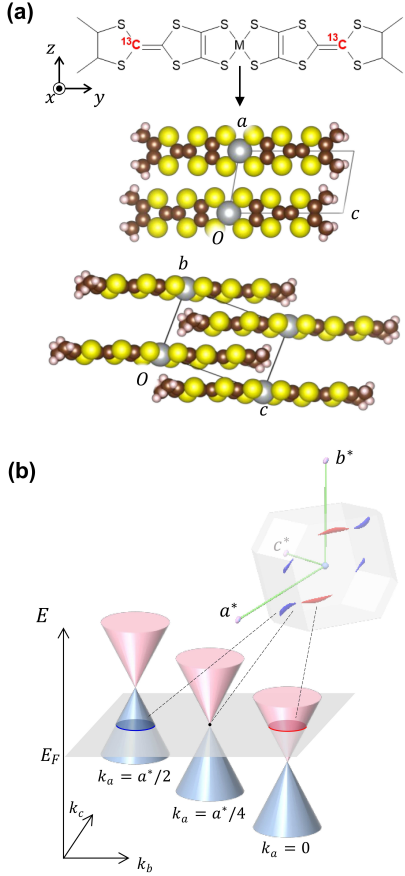} 
\label{crystal}
\caption{(a) Molecular structure of Ni(dmdt)$_2$ and crystal structure of [Ni(dmdt)$_2$]. The carbon cites enriched by $^{13}$C isotopes to 99\% are colored red. The unit cell contains only one Ni(dmdt)$_2$ molecule and the two $^{13}$C isotopes shown are mutually related with inversion symmetry; thus, all $^{13}$C isotopes are crystallographically equivalent. (b) Hole-like (blue) and electron-like (red) Fermi pockets along DNLs in the 3D Brillouin zone of \nidmdt obtained by first-principles calculations \cite{Kobayashi2021}. The Dirac points in the $k_b$--$k_c$ plane undulate around $E_{\rm F}$ along the $a^*$ direction, forming small electron (hole) pockets. The Dirac cones are depicted to be symmetric for simplicity although tilted in this compound.}
\end{figure}

The present study aims to microscopically verify the DNLs in \nidmdt and elucidate the correlation and magnetic-field effect in it by $^{13}$C-NMR spectroscopy combined with numerical simulations.
We performed $^{13}$C-NMR experiments on a polycrystalline sample of $^{13}$C-enriched [Ni(dmdt)2] (Fig. 1(a)) under a magnetic field of 11 T. The NMR spin echo signals following the $(\pi/2)_x$--$(\pi)_x$ pulse sequence were Fourier-transformed into NMR spectra. The nuclear spin-lattice relaxation rate, 1/$T_1$, was measured by the saturation recovery method. The relaxation of the nuclear magnetization, $M(t)$, was not single-exponential in time at all temperatures because the anisotropic hyperfine coupling constants at the $^{13}$C sites give a distribution of 1/$T_1$ in a randomly oriented polycrystal against the applied field. Then, the volume average of 1/$T_1$ was determined by the initial slope of the relaxation curve, which is denoted by 1/$^iT_1$.

To get physical insight into the experimental results, we also performed numerical simulations of the Knight shift, $K$, and 1$/^iT_1T$ in the-density-of-state (DOS) approximation for DNLs. Generally, $K$ is related to the static spin susceptibility, $\chi_{\mathrm{s}}$, by $K = A_{\parallel} \chi_{\mathrm{s}}$, where $A_{\parallel}$ is the parallel component of the hyperfine coupling constant to the magnetic field $\bm{B}_0 = B_0(\cos \phi \sin \theta, \sin \phi \sin \theta, \cos \theta)$ with $\phi$ and $\theta$ defined as angles from the $x$ axis in the $xy$ plane and from the $z$ axis, respectively, and $B_0=11 \mathrm{T}$. Here, $x$, $y$, and $z$ denote the molecular principal axes shown in Fig. 1(a). Using $\chi_{\mathrm{s}}=\mu_\mathrm{B}\int dE D(E)(-\partial f/\partial E)$ with $D(E)$, the spinless DOS and $f(E)$, the Fermi--Dirac distribution function with chemical potential fixed to the charge-neutrality point, the polycrystalline average of the isotropic term of the Knight shift, $K_\mathrm{iso}$, is expressed as
\begin{equation}
K_{\mathrm{iso}}=\mu_\mathrm{B} \iint \frac{d\phi d\theta \sin \theta}{4\pi} A_{\parallel}(\theta, \phi) \int dE D(E) \left(-\frac{\partial f}{\partial E} \right). \label{KnightShift}
\end{equation}
where $\mu_{\mathrm{B}}$ is the Bohr magneton, and $A_\parallel (\theta,\phi)=A_{xx}  \cos^2\phi \sin^2\theta +A_{yy} \sin^2\phi \sin^2\theta+A_{zz} \cos^2\theta$ with the principal values of the hyperfine coupling tensor, $A_{xx}$, $A_{yy}$ and $A_{zz}$. The polycrystalline average of 1/$T_1T$, \ToneTbar, in the DOS approximation yields \cite{Kawamoto1995}
\begin{eqnarray}
\overline{\frac{1}{T_1T}}= \frac{\pi k_{\mathrm{B}}}{\hbar} \left( \gamma_{\rm e} \gamma_{\rm n} \hbar^2 \right)^2\iint \frac{d\phi d\theta \sin \theta }{4\pi}A_{\perp}^2(\theta, \phi) \notag \\
 \times \int dE \left\{ D(E)\right\}^2 \left( -\frac{\partial f}{\partial E} \right), \label{T1T}
\end{eqnarray}
where $A_{\perp}$ is the transverse component of the hyperfine coupling constant against the magnetic field $\bm{B}_0$, $\gamma_{\rm e}$ ($\gamma_{\rm e}$ ) is the electron ($^{13}$C nuclear) gyromagnetic ratio, $\hbar$ is the reduced Planck constant, $k_{\rm B}$ is the Boltzmann constant, and $A_{\perp}^2(\theta ,\phi)= (A_{yy}^2+A_{zz}^2 )  \cos^2\phi  \sin^2\theta + (A_{zz}^2+A_{xx}^2)  \sin^2\phi \sin^2\theta +(A_{xx}^2+A_{yy}^2 ) \cos^2\theta $. As the $^{13}$C hyperfine coupling tensor is mainly determined by the on-site $p_z$  orbital, it is reasonably assumed to be uniaxial; namely, $A_{xx}=A_{\rm iso}+2A_{\rm aniso}$ and $A_{yy}=A_{zz}=A_{\rm iso}-A_{\rm aniso}$ with the isotropic (anisotropic) term, $A_{\rm iso(aniso)}$. In the simulations, we used the values of $A_{\rm iso(aniso)} = 3.9 (3.0) \rm \ kOe/(\mu_{\mathrm{B}}\ dmdt)$ for the analogous material [Ni(tmdt)$_2$] with the similar molecular flamework \cite{Takagi2016} because the $A_{\rm iso(aniso)}$ values for \nidmdt are not available.

We simulated $K$ and \ToneTbar \ of a DNL model with the Fermi velocity of 2D Dirac cones on the $k_b$--$k_c$ plane, \vF, and the transfer integral along the $a$-axis, $t_a$, as illustrated in Fig. 2(a). First, we employed a zero-field (ZF) model, which ignores the effect of Landau quantization. The energy band is given as
\begin{equation}
E^{\rm ZF}_{\bm{k}} = \hbar v_{\rm F} \sqrt{k_b^2 + k_c^2} + 2t_a \cos ak_a. \label{E_ZF}
\end{equation}
The DOS, $\sum_{\bm{k}}\delta(E-E_{\bm{k}})$, is expressed as
\begin{equation}
D^{\rm ZF}( E) = \dfrac{|E|}{2a\pi (\hbar v_{\rm F})^2}\qquad \qquad \left( |E| \geq 2t_a \right) \label{DOS_ZF1} \\
\end{equation}
and
\begin{align}
D^{\rm ZF}(E) &= \dfrac{|E|\left(\pi - 2 \cos ^{-1} [E/2t_a]\right)+4t_a \sqrt{1-(E/2t_a)^2}}{2a\pi^2 \left( \hbar v_{\rm F}\right)^2}\notag \\
&\hspace{4.6cm}\left( |E| < 2t_a\right). \label{DOS_ZF2}
\end{align}
As seen in Fig. 2(b), $D^{\rm ZF}(E)$ is linear to $|E|$, as in 2D DFs, for $|E|>2t_a$, and is quadratic-like for $|E|<2t_a$ with a finite value at $E=0$ due to the Fermi pockets arising from the dispersive DNLs.

\begin{figure}[h]
\centering
\includegraphics[width = \columnwidth]{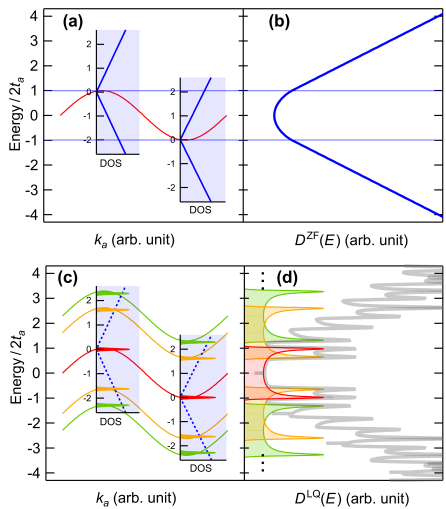} 
\label{DOSsimulation}
\caption{Dispersion and density of states of DNL with and without Landau quantization used in the simulations. (a) $k_a$ dispersion of the 2D DF DOS and (b) the whole DOS in the DNLs under zero field. (c) $k_a$ dispersions of the Landau levels in the 2D plane and (d) the DOS of the DNL under a finite field. In (d), the DOS of each Landau level is distinguished by different colors and the whole DOS is shown by a gray line.}
\end{figure}

Next, we performed the simulations incorporating the Landau-quantization (LQ) effect. In 2D DF systems under a magnetic field $\bm{B}$, the Landau levels appear at $E_{\pm N}=\pm v_{\rm F} \sqrt{2e\hbar B|N|}$ with an integer index $N$~\cite{CastroNeto2009} and have the dispersions given by
\begin{equation}
E^{\rm LQ}_N (k_a,B_{\perp}) = sgn(N)v_{\rm F}\sqrt{2e\hbar B_{\perp}|N|} + 2t_a\cos ak_a, \label{E_LQ}
\end{equation}
where $sgn(x)$ is the sign function, and $B_{\perp}$ is a field component normal to the $k_b$--$k_c$  plane (parallel to the $a$-axis). Each Landau mode of the 2D DFs in the $k_b$--$k_c$  plane maintains the dispersion along the $a$-axis even under magnetic field and forms a 1D band along the $a$-axis (Fig. 2(c)). We assume that the crystal $a$-axis is parallel to the molecular $z$-axis (Fig. 1(a)), although off by less than $20^{\circ}$ from parallel, to simplify the integration over $\theta$ and $\phi$ in Eqs. (\ref{KnightShift}) and (\ref{T1T}); namely, $B_{\perp}=B_0  \cos\theta$. The DOS in the LQ model is given as
\begin{equation}
D^{\rm LQ}(E, B_\perp) = \frac{eB_\perp}{2\pi \hbar }\sum_{N, k_a}\delta  \left(E-E^{\rm LQ}_N(k_a, B_\perp)\right). \label{DOS_LQ}
\end{equation}

In calculating $K_{\rm iso}$ and \ToneTbar, $D(E)$ in Eqs. (\ref{KnightShift}) and (\ref{T1T}) is replaced by $D^{\rm ZF} (E)$ (Eqs. (\ref{DOS_ZF1}) and (\ref{DOS_ZF2})) or $D^{\rm LQ} (E,B_\perp)$ (Eq.~(\ref{DOS_LQ})). As seen in Fig.~2(d), $D^{\rm LQ} (E,B_\perp)$ is divergent at $E=E_N\pm 2t_a$, the edges of the 1D $N$-th Landau-level band. In calculating \ToneTbar, we replaced the delta function in Eq. (\ref{DOS_LQ}) with the Gaussian with a width of $k_{\rm B}$ to avoid divergence in the integral.

Figure 3(a) displays the temperature variation of $^{13}$C-NMR spectra, which become somewhat broadened and shifted in the negative direction upon cooling. The first moment of the spectra, which represents the isotropic component of the NMR shift tensor, varies linearly with temperature in the range of 40--300 K (Fig. 3(b)), as expected in DFs with DOS that has linear dependence on energy. The linewidth characterized by the square root of the second moment of the spectra appreciably increases below100 K (inset of Fig. 3(b)). Similar behavior is observed in [Zn(tmdt)$_2$], in which $\chi_{\mathrm{s}}$ increases with $T$ due to triplet excitations from a singlet ground state \cite{Takagi2017}, and attributed to different $T$-dependences of the principal values of the NMR shift tensor comprised of chemical shift and $T$-dependent spin shift. This is likely the case in the present system as well.

\begin{figure*}
\centering
\includegraphics[width = 17.2cm]{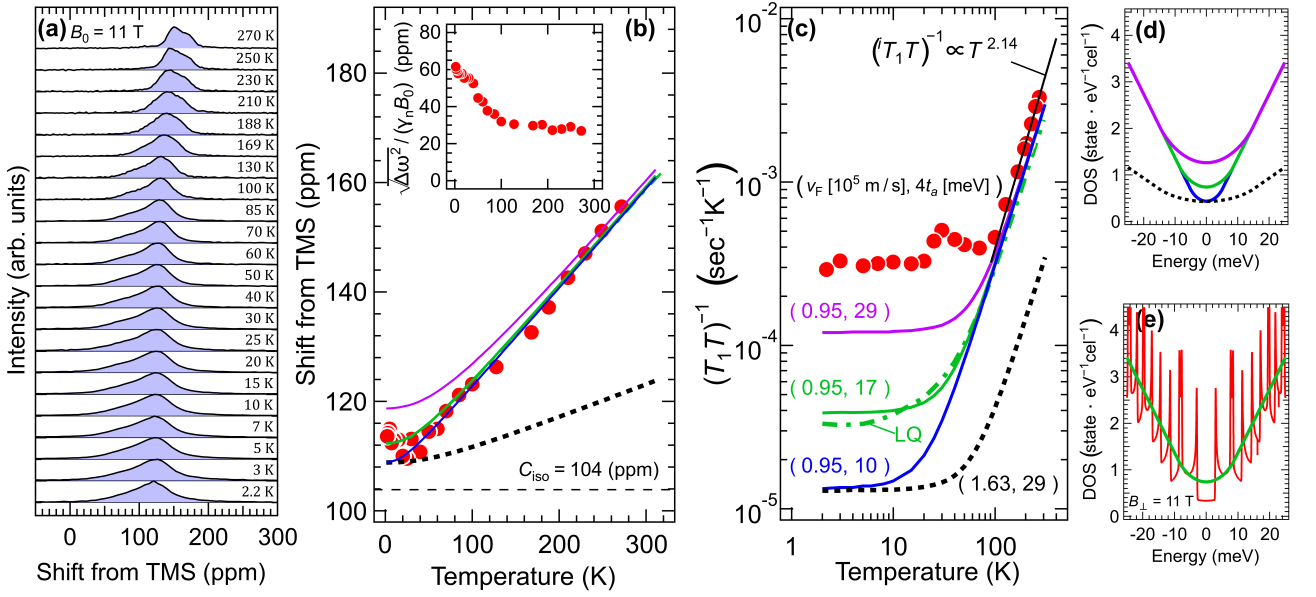}
\label{results}
\caption{(a) $^{13}$C-NMR spectra of polycrystalline \nidmdt under 11 Tesla. The origin of the spectral shift is the spectral position of tetramethylsilane (TMS). (b) First moment of $^{13}$C-NMR spectra (red circle), the ZF-model simulations with $\left(v_{\rm F},4t_a \right)=(0.95\times10^5 \rm~m/s, 29~meV)$(purple), $(0.95\times10^5 \rm~m/sec, 17~meV)$(green), $(0.95\times10^5 \rm~m/sec, 10 ~meV)$(blue) and $(1.63\times10^5 \rm~m/sec, 29~meV)$(black dash line) and the LQ-model simulation with $\left(v_{\rm F},4t_a \right)=(0.95\times10^5 \rm~m/sec, 17~meV)$ (green dotted line) Isotropic term of chemical shift, $C_{\rm iso}$, is taken to 104 ppm (see text). Inset shows the square root of the second moment of the observed spectra. (c) $^{13}$C nuclear spin-lattice relaxation rate divided by temperature, 1/$T_1T$; experimental data (red square) and simulations (the same symbols as (b)). In experiments, the volume-averaged 1/$T_1T$ for a polycrystal is determined from the initial decay rate of the observed nuclear magnetization (red square) and, in simulations, 1/$T_1T$ is angle-averaged (the same symbols as (b)) to be compared with the experiments. (d) DOS of the ZF model. (e) DOS of the LQ model with the same parameters as (b) and (c) and a field component normal to the $k_b$--$k_c$ plane, $B_0 \cos \theta= 6.00$ T (red line). Green line is the DOS of the ZF model.}
\end{figure*}

Figure 3(c) shows the temperature dependence of \ToneTini. Above 100 K, \ToneTini is proportional to $T^{2.14}$ with an exponent slightly larger than 2, an exponent expected in the linear-in-$T$ DOS. The NMR study of \alphaithree found that the Dirac cone is sharpened around the Dirac points due to the velocity renormalization of the long-ranged Coulomb interaction that is insufficiently screened in DF systems \cite{Hirata2021}. This Dirac cone reshaping makes the energy dependence of DOS superlinear, which lifts the temperature-exponent to above 2, giving an explanation to the observed exponent. Only a slight deviation from 2 and the nearly $T$-linear Knight shift suggest that screening works  to some extent due to the small Fermi pockets.

On cooling below 100 K, \ToneTini saturates as in conventional metals. This Korringa relation is an indication of small Fermi pockets at $E_{\rm F}$. The energy scale of the crossover from the metal to DF regimes, $\sim$100 K, is a few times smaller than the half of the dispersion width of the DNL, $\sim$15 meV, predicted by first-principles calculations, as is discussed in comparison with the simulation. Before entering the metallic regime at low temperature, \ToneTini shows a subtle peak at around 30 K. As seen below, this feature is unexplainable in terms of the DOS model of DNLs.

The ZF-simulation results for $K_{\rm iso}$ and  \ToneTbar are shown in Figs. 3(b) and 3(c), respectively, for several values of \vF and $4t_a$. $K_{\rm iso}$ and \ToneTbar follow the 2D DF-like $T$-dependence ($K\propto T,~1/T_1T\propto T^2$) at high temperatures and both level off below the crossover temperature $T^{\rm cr}\equiv 2t_a/k_{\rm B}$, consistent with the DOS profile of the DNLs as displayed in Fig. 2(b). In Figs. 3(b)--(d), the black dash line shows the simulation with the angle-averaged \vF value of first principles calculations, $\overline{v_{\rm F}} =1.63\times 10^5 \rm~m/sec$. Obviously, there is a large gap between the experimental data and the simulations. From the slope of $K_{\rm iso}$ vs $T$ in $T>T^{\rm cr}$, which is proportional to $v_{\rm F}^{-2}$, the real \vF value in \nidmdt is estimated at $v_{\rm F}=0.95\times10^5 \rm~m/sec$, which is 1.7 times reduced from the theoretical value $\overline{v_{\rm F}}$. Thus, the band width, which is proportional to \vF, is reduced by a factor of 1.7 from the first-principle value very probably due to the short-range part of the Coulomb interactions \cite{Hirata2021,Hirata2016}.

Then, with the \vF value fixed at $0.95\times 10^5$ m/sec, we varied the dispersion width, $4t_a$, which determines the DOS in $|E|<2t_a$ (Figs. 2(b) and 3(d)). The residual DOS, $D^{\rm ZF} (E=0)$, gives the $T$-independent values of \Kiso  and \ToneTbar below $T^{\rm cr}$, both of which increase with $t_a$. The simulated \Kiso with $4t_a=10,17$ and 29 meV is shown by curves in Fig. 3(b). The $4t_a$ values in $10<4t_a<17$ meV in conjunction with the chemical shift $C_{\rm iso}=104$ ppm reproduce the experimental behavior, suggesting that the band width is 1.7--3 times narrower than the first-principles value of the dispersion width, $\Delta E=29$ meV. The ratio $v_{\rm F}/\overline{v_{\rm F}} = 0.58$ is nearly equal to $4t_a/\Delta E= 0.59$ when $4t_a=17$ meV, consistently indicating a reduction in the band width. The $C_{\rm iso}$ value is reasonably close to $C_{\rm iso}$ of 126 and 114 ppm, respectively, in [Zn(tmdt)$_2$] and [Au(tmdt)$_2$] with tmdt ligand of common molecular frame to dmdt \cite{Takagi2017, Takagi2020}.

\ToneTbar in the DF regimes above $T^{\rm cr}$ is nearly solely determined by $v_{\rm F}$. As shown in Fig. 3(c), the first-principles value, $v_{\rm F}=1.63\times10^5$ m/sec, fails to explain the experiments. However, the value of $v_{\rm F}=0.95\times10^5$ m/sec determined from the Knight shift gives an excellent agreement irrespectively of the choice of the $4t_a$ value. The ratio of the experimental \ToneTini value to the simulated one, e.g., at 200 K is 1.3, which points to moderate antiferromagnetic spin correlations above $T^{\rm cr}$. At low temperatures below $T^{\rm cr}$, \ToneTbar is determined by the residual DOS that depends on $4t_a$. As $4t_a$ is increased, the level of \ToneTini rises like the Knight shift but stays considerably below the experimental values for all the values of $4t_a$ (Fig. 3(c)).
With $4t_a = 17~\mathrm{meV}$, the ratio of \ToneTini to $\overline{1/T_1 T}$ is approximately 8.0 below 5 K. This feature is regarded as an indication of enhanced antiferromagnetic spin correlations in the Fermi pockets. It can be accounted for by the nesting instability between electron pockets and hole pockets. On the other hand, the enhancement factor of the relaxation rate is considerably reduced above $T^{\mathrm{cr}}$. This suppression of the correlation effect at high temperatures is reasonably due to the progressive screening of the Coulomb interactions by an increasing number of quasiparticles at elevated temperatures.
In our simulations, there was no indication of a peak in \ToneTbar as observed in experiments.

The possible effect of the Landau quantization on \Kiso and \ToneTbar was examined with the LQ model with $\left(v_{\rm F}, 4t_a\right) = (0.95\times10^5  \rm~m/sec,17~meV)$. As indicated by a green dotted line in Fig. 3(b), \Kiso of the LQ model is indistinguishable from that of the ZF model in entire temperatures. Regarding \ToneTbar, there is no appreciable difference between the two models above 100 K because the $4t_a$-dispersed Landau levels overlap in the energy region well above $E_{\rm F}$ as depicted in Fig. 2(d).
In the 10--70 K temperature range, \ToneTbar of the LQ model is slightly larger than that of the ZF model. The reason why this \ToneTbar is more sensitive to the Landau quantization than \Kiso is very probably that the divergent parts of DOS at the edges (Fig. 3(e)) are more influential to its square $( \propto \overline{1/T_1 T})$ than to itself $\left( \propto K_{\rm iso}\right)$.
On the other hand, \ToneTbar of the LQ model below 10 K becomes smaller. 
$D^{\rm  LQ}(E)$ at $E_{\rm F}$ is determined only by the $N=0$ Landau level when $E^{\rm LQ}_{N \geq 1}(k_a, B_\perp) > 0$ with any values of $k_a$ (Eq. (\ref{E_LQ})). 
A lower limit of $B_\perp$ for $E^{\rm LQ}_{N \geq 1}(k_a, B_\perp) > 0$ is approximately 6 T with $\left(v_{\rm F}, 4t_a\right) = (0.95\times10^5  \rm~m/sec,17~meV)$.
Thus, under the field of $B_\perp > 6~$T, $D^{\rm  LQ}(E)$ in the vicinity of $E_{\rm F}$ is smaller than $D^{\rm  ZF}(E)$  as shown in Fig. 3(e).
Such suppression of $D^{\rm  LQ}(E)$ affects the reduction of $\overline{1/T_1 T}$ of the LQ model in the low temperature region.

We note that the LQ model also does not explain the subtle peak around 30 K. A similar peak was observed in $^{125}$Te-NMR 1/$T_1T$ of ZrSiTe \cite{Tian2021}, which is considered to originate in the Lifshitz transition that occurs when the chemical potential pass through the point of the van Hove singularity. The present system, however, has no such van Hove singularities under zero field and, even under magnetic field, the edge singularities of Landau-level bands are smeared out in randomly oriented grains against the field direction. Recently, Kawamura and Kobayashi have studied the multi-orbital Hubbard model of DNL and suggested the enhancement in 1/$T_1T$ due to intramolecular antiferromagnetic fluctuations \cite{Kawamura2022}. The small peak in question may be its symptom.

In summary, we conducted $^{13}$C-NMR experiments on a polycrystalline sample of the single-component molecular material, \nidmdt, suggested as a DNL system by the first-principles calculations and discussed the experimental results in the light of model simulations with and without the Landau quantization effect. The $T$-linear Knight shift and $T$-quadratic \ToneTini that level off at low temperatures evidence slightly dispersive DNLs with small Fermi pockets. The Knight shift and \ToneTini data compared with the simulations found that the dispersion width, $4t_a$ and the Fermi velocity, $v_{\rm F}$, are both reduced from the first-principles values by a factor of 1.7--3, indicative of a band narrowing, and that antiferromagnetic spin correlations are highly enhanced in the low-temperature metallic state. In addition, the simulation projects that the Landau quantization causes a slight change in the NMR relaxation rate at low temperatures but makes no influence in the Knight shift. All these results verify the Dirac-nodal-line picture of \nidmdt and reveal the correlation effect on the band structure and the energy-profile of spin fluctuations.

The authors thank Akito Kobayashi and Taiki Kawamura for fruitful discussions. This work was supported by the JSPS Grants-in-Aid for Scientific Research (Grant Nos. 17K05846, 18H05225, 19H01846, 20K20894, 20KK0060 and 21K18144), the Mitsubishi Foundation (Grant No. 202110014), and the Japan Science
and Technology Support for Pioneering Research Initiated
by the Next Generation (Grant No. JPMJSP2108). We also thank the Cryogenic Research Center at the University of Tokyo for supporting the low-temperature experiments.

\end{document}